\documentclass[12pt,preprint]{aastex}

\slugcomment{Submitted to Astronomical Journal}

\shorttitle{IDV in blazars}
\shortauthors{Gupta et al.}

\begin{document}


\title{Optical Intra-day Variability in Blazars}  
%


\author{Alok C. Gupta\altaffilmark{1}, J. H. Fan\altaffilmark{2}, 
J. M. Bai\altaffilmark{1} and S. J. Wagner\altaffilmark{3}}

\email{acgupta30@gmail.com}
%
%

%
\altaffiltext{1}{National Astronomical Observatories/Yunnan Observatory, Chinese
Academy of \\ 
\hspace*{0.22in} Sciences, P.O. Box 110, Kunming, Yunnan 650011, China.}
\altaffiltext{2}{Center for Astrophysics, Guangzhou University, Guangzhou 510006, China}
\altaffiltext{3}{Landessternwarte, K$\ddot{o}$nigstuhl 12, D - 69117, Heidelberg, Germany}
%

\begin{abstract}
We selected a sample of a dozen blazars which are the prime candidates for 
simultaneous multi-wavelength observing campaigns in their outburst phase. We searched 
for optical outbursts, intra-day variability and short term variability in these blazars. 
We carried out optical photometric monitoring of nine of these blazars in 13 observing 
nights during our observing run October 27, 2006 $-$ March 20, 2007 by using the 1.02 
meter optical telescope equipped with CCD detector and BVRI Johnson broad band filters 
at Yunnan astronomical observatory, Kunming, China. From our observations, 
our data favor the hypothesis that
three blazars: AO 0235$+$164, S5 0716$+$714 and 3C 279 were in the outburst state; 
one blazar: 3C 454.3 was in the post outburst state; three blazars: S2 0109$+$224, 
PKS 0735$+$178 and OJ 287 were in the pre/post outburst state; one blazar: ON 231 was 
in the low-state; and the state of one blazar: 1ES 2344$+$514 was not known because 
there is not much optical data available for the blazar to compare with our observations.
We observed densely sampled 1534 image frames of these nine blazars. Out of three nights 
of observations of AO 0235$+$164, intra-day variability was detected in two nights. Out 
of five nights of observations of S5 0716$+$714, intra-day variability was detected in 
two nights. In one night of observations of PKS 0735$+$178, intra-day variability was 
detected. Out of six nights of observations of 3C 454.3, intra-day variability was 
detected in three nights. No intra-day variability was detected in S2 0109$+$224, OJ 287,
ON 231, 3C 279 and 1ES 2344$+$514 in their 1, 4, 1, 2 and 1 nights of observations 
respectively. AO 0235$+$164, S5 0716$+$714, OJ 287, 3C 279 and 3C 454.3 were observed in
more than one night and short term variations in all these blazars were also noticed.
From our observations and the available data, we found that 
the predicted optical outburst with the time interval of $\sim$ 8 years in AO 0235$+$164 
and $\sim$ 3 years in S5 0716$+$714 have possibly occurred. 

\end{abstract}





\section{Introduction}

Blazars represent a small subset of the most enigmatic class of radio-loud active 
galactic nuclei (AGN), exhibiting strong variability at all wavelengths of the whole 
electromagnetic (EM) spectrum, strong polarization from radio to optical wavelengths, 
and usually core dominated radio structures. The radiation of blazars at all wavelengths 
is predominantly nonthermal. This class includes BL Lacertae (BL Lac) objects and 
flat-spectrum radio quasars (FSRQs). BL Lacs show largely featureless optical continuum. 
In a unified model of the radio-loud AGN based on the angle between the line of sight 
and the emitted jet from the source, blazars jet make angle of $\sim$ 10$^{\circ}$ from 
the line of sight (Urry \& Padovani 1995). The radiation emitted by the plasma, with bulk 
relativistic motion in the jet oriented at small viewing angles, is affected by relativistic 
beaming, which in turn implies a shortening of time scales by a factor $\delta^{-1}$, where 
$\delta$ is the Doppler factor. 

From observations of blazars, it is known that they vary on the diverse time scales. 
Variability time scales of blazars can be broadly divided into 3 classes viz. 
intra-day variability (IDV) or micro-variability, short term outbursts and long term 
trends. Significant variations in flux of a few tenths of magnitude over the course 
of a day or less is often known as IDV (Wagner \& Witzel 1995). Short term outburst 
and long term trends can have time scales range from few weeks to several months and 
several months to years, respectively. In last about two decades, variability of 
blazars in radio to optical bands on diverse time scales have been reported in a 
large number of papers (e.g. Miller et al. 1989; Courvoisier, et al. 1995; Heidt \& 
Wagner 1996; Takalo et al. 1996; Sillanp$\ddot{a}\ddot{a}$ et al. 1996a, 1996b; 
Bai, et al. 1998, 1999; Fan et al. 1998, 2002, 2007; Xie, et al. 2002a; Gupta et al. 
2004; Ciprini et al. 2003, 2007 and references therein). 

We selected a sample of a dozen blazars which are prominent candidates for 
simultaneous multi-wavelength observing campaigns in their outburst phase. The 
motivation of the present work was to observe these blazars in search for IDV, short 
term variability and also find out if there is any one being in the outburst state. 
Blazar emission mechanism in the outburst state and detected IDV is strongly 
supported by the jet based models of radio-loud AGN. In general, blazar emission 
in the outburst state is nonthermal Doppler boosted emission from jets (Blandford 
\& Rees 1978; Marscher \& Gear 1985; Marscher et al. 1992, Hughes et al. 1992). 
There are other models of AGN that can explain the IDV in any type of AGN are optical 
flares, disturbances or hot spots on the accretion disk surrounding the black hole of 
the AGN (Mangalam \& Wiita 1993 and references therein). Models based on the instabilities 
on accretion disc are mainly supported blazars IDV when the blazar is in the low-state. 
When a blazar is in the low-state, any contribution from the jets if at all present, is 
very weak. Recently, it is noticed that, in the low luminosity AGN, accretion disk is 
radiatively inefficient (Chiaberge et al. 2006; Capetti et al. 2007). So, there will 
be an alternative way to explain the IDV in the low-state of blazars, in which a weak 
jet emission will be responsible for the IDV. With this motivation we recently carried 
out optical photometric observations of the nine blazars: S2 0109$+$224, AO 0235$+$164, 
S5 0716$+$714, PKS 0735$+$178, OJ 287, ON 231, 3C 279 and 1ES 2344$+$514 in R passband 
and 3C 454.3 in V and R passbands.  

The paper is arranged as follows: section 2 describes observations and data analysis
method, in section 3 we mentioned our results, discussion and conclusion of the present 
work is reported in section 4. 

\section{Observations and Data Reductions}

The photometric observations of the blazars S2 0109$+$224, AO 0235$+$164, S5 0716$+$714, 
PKS 0735$+$178, OJ 287, ON 231, 3C 279 and 1ES 2344$+$514 in Johnson R passband and 
3C 454.3 in Johnson V and R passbands were carried out using the CCD detector (1024 pixels 
$\times$ 1024 pixels) mounted at f/13.3 Cassegrain focus of the 1.02 meter RC (Ritchey 
Chertien) optical telescope at Yunnan observatory, Kunming, China. These blazars were 
observed during October 27, 2006 to March 20, 2007 in 13 observing nights. Each pixel 
of the CCD detector projected on the sky corresponds to 0.38 arcsec in both the dimensions. 
The entire CCD chip covers $\sim$ 6.5 $\times$ 6.5 arcmin$^{2}$ of the sky. The read out 
noise and gain of the CCD were 3.9 electrons and 4.0 electrons/ADU respectively. Throughout 
the observing run, the typical seeing was $\sim$ 1.8 arcsec ranging from 1.5 to 2.0 arcsec. 
Several bias frames were taken intermittently in each observing night and twilight sky flats 
were taken in V and R passbands. Detail photometric observation log is given in the Table 1.

For each observing night we generated a master bias by taking median of all bias frames 
in the night. Master bias frame is subtracted from all the target image frames and flat 
field image frames of the night. Then master V and R passband flat field image frames 
were generated for the night by taking median of all flat field images in V and R passband 
respectively. Image processing or pre processing (bias subtraction, flat-fielding and 
cosmic rays removal) were done using standard routines in IRAF\footnote{IRAF is distributed 
by the National Optical Astronomy Observatories, which are operated by the Association of 
Universities for Research in Astronomy, Inc., under cooperative agreement with the National 
Science Foundation.} (Image Reduction and Analysis Facility) software. Photometric reduction 
or processing of the data were performed by aperture photometric technique using DAOPHOT II 
(Dominian Astronomical Observatory Photometry) software (Stetson 1987). Aperture photometry 
was carried out with concentric apertures of radii 8, 11, 15 and 20 pixels. The data reduced 
with different aperture radii were found to be in good agreement. However, it was noticed 
that the best signal to noise ratio was obtained with the aperture radius of 15 pixels (5.7 
arcsec slightly more than 3 $\times$ of typical FWHM). 

In our observations of all nine blazars, local standard stars were also always present in 
the observed image frames. We always observed three or more local standard stars in our every 
target blazar field. The magnitude of standard stars which we observed in the field of our 
observed blazars are given in Table 2. These standard stars were used to check the non variable 
characteristic of standard stars and then transformed the instrumental magnitude of the 
blazars to the standard magnitude. Finally we selected two standard stars from each blazar 
field and plotted their differential instrumental magnitude light curve by adding arbitrary 
offset for clarity in the plot with the light curves of blazars (calibrated magnitude). 
Photometric data of our blazars observations are provided in 
Table 4\footnote{Here we provide photometric data for the blazars S2 0109$+$224, S5 0716$+$714,
PKS 0735$+$178, ON 231 and 1ES 2344$+$514. Data of the blazars AO 0235$+$164, OJ 287, 3C 279
and 3C 454.3 will be stored in the WEBT (http://www.to.astro.it/blazars/webt) archive. For 
questions regarding their avalability, please contact the WEBT President Massimo Villata 
(villata@oata.inaf.it)}.

\section{Results}

\subsection{Variability Detection Criterion}

Using aperture photometry of blazar and few standard stars in the blazar field, we determined 
the differential instrumental magnitude of blazar $-$ standard star A, blazar $-$ standard star B
and standard star A $-$ standard star B. Finally we selected star A or B to do calibration of the
blazar data. We determined observational 
scatter from blazar $-$ standard star A $\sigma$ (BL $-$ Star A), blazar $-$ standard star B 
$\sigma$ (BL $-$ Star B) and standard star A $-$ standard star B $\sigma$ (Star A $-$ Star B). 
The variability of target blazar is investigated by using variability parameter C, introduced 
by Romero et al. (1999).

The variability parameter C is expressed as the average of C$_{1}$ and C$_{2}$ 
\begin{eqnarray}   
C_{1} = \frac {\sigma (BL - Star A)}{\sigma (Star A - Star B)} \hspace*{0.2in} \rm{and} \hspace*{0.2in} 
C_{2} = \frac {\sigma (BL - Star B)}{\sigma (Star A - Star B)}
\end{eqnarray}
If C $>$ 2.57, the confidence limit of variability is 99\%. 
The value of C by using the two standard stars for all the nine blazars for different observing 
nights are reported in Table 3.  

Intra-day variability amplitude defined as (Heidt \& Wagner 1996)
\begin{eqnarray}
A = 100 \times \sqrt {(A_{max} - A_{min})^{2} - 2\sigma^{2}} \hspace*{0.1in}\%
\end{eqnarray}

where A$_{max}$ and A$_{min}$ are the maximum and minimum magnitude in the 
calibrated light curve of the blazar. $\sigma$ is the averaged measurement error of the
blazar light curve. 

\subsection{IDV and Short Term Variability of Individual Blazars}

\noindent 
{\bf S2 0109$+$224}

S2 0109$+$224 is a BL Lac object which has featureless optical spectrum. The host galaxy 
of the source was unresolved in NTT observations (Falomo 1996) and also in K passband 
observations using UKIRT (Wright et al. 1998). Falomo (1996) suggested that the redshift 
of the blazar z $\geq$ 0.4 based on its optical appearance, and assuming M$_{R} = -$23.5 
for it, a value similar to that characterizing some galaxies located at few arcseconds 
north-east of the object. The blazar has shown optical flux 
variability on diverse time scales 
(Ciprini et al. 2003). The degree of optical polarization varied between 10\% to 30\% 
(Takalo 1991; Valtaoja et al. 1993). 

We observed S2 0109$+$224 on January 11, 2007 in R passband. The light curves of the blazar 
and differential instrumental magnitude (star D  - star C1) with arbitrary offset are 
displayed in the bottom panel of Fig. 6. From the figure it is clear that the source has not 
shown IDV on one night of our observations. We also performed the IDV detection test described 
in the variability detection criterion section and found that the value of C is 1.07, which 
also confirms the non detection of IDV in our observations. Ciprini et al. (2003) have done 
the multi band optical monitoring of the source for seven years (1994 $-$ 2002) and also 
plotted its B passband historical light curve (1906 $-$ 2002). 1994 $-$ 2002 light curves of 
the source show the brightest and faintest states of the source in R passband $\sim$ 13.2 and 
$\sim$ 16.5 magnitudes. In the present observations, we observed the source $\approx$ 1.0 
magnitude fainter than its brightest state and $\approx$ 2.3 magnitude brighter than its 
faintest state. 
Therefore, it appears likely that we have observed the source in its pre/post outburst state.  

\noindent
{\bf AO 0235$+$164}

AO 0235$+$164 at (redshift z = 0.94) is classified as a BL Lac object by Spinrad \&
Smith (1975). This blazar has been extensively observed in the recent past from
radio to X-ray bands and has shown variations in all those bands on diverse time scales
(Ghosh \& Soundararajaperumal 1995; Heidt \& Wagner 1996; Romero
et al. 1997, 2000; Webb et al. 2000; Raiteri et al. 2001; Padovani et al. 2004; and
references therein). It is one of the blazars which has shown very high polarization
P$_{V} =$ 44\% and P$_{IR} =$ 36.4\% (Impey et al. 1982; Stickel et al. 1993; Fan \& Lin
1999). Using 25 years (1975 - 2000) of radio and optical data, Raiteri et al. (2001) 
predicated that the blazar should show a possible correlated periodic radio and optical 
outburst on the time scale of 5.7$\pm$0.5 years that was expected 
in February - March 2004. An intense WEBT (Whole Earth Blazar Telescope) multi-wavelength 
observing campaign during 2003$-$2005 was organized and the predicted out burst could not 
be detected (Raiteri et al. 2005; 2006a; 2006b). Raiteri et al. (2006b) did 
analysis of the long term optical light curves and reported that the optical outburst may 
have a longer time scale $\sim$ 8 years.

We observed AO 0235$+$164 on January 09 and 10, and March 19, 2007 in R passband.
The light curve of the blazar and differential instrumental magnitude (star 1 - star 3)
with an arbitrary offset are displayed in Fig. 1. From the figure it is seen that the source 
has shown IDV on January 10 and March 19, 2007. We also performed the IDV detection test and 
found that the value of C for January 09, January 10 and March 19 are 2.00, 4.14 and 3.73 
respectively which confirms that the source has shown IDV in the observations of January 10 and 
March 19. The variability amplitude on January 10 and March 19 are 13.7\% and 9.5\% respectively. 
Day to day variations are distinctly visible in Figs. 1 and 7. In Fig. 7 the filled circles show 
the day average magnitude of AO 0235$+$164 and open circles show the day average of differential 
magnitude of two standard stars (star 1 and star 3) offset by the same arbitrary constant on all 
the three observing nights. Fig. 7 also shows the variability trend and if the source has reached 
even brighter state than January 10, it will be before March 19, 2007. From variability trend from 
observations show the source was still in outburst phase on March 19. Most interestingly, if we 
compare our Figs. 1 and 7 to the long term light curve (1990 - 2005) of (Fig. 4 top panel of 
Raiteri et al. 2006b), we found that we have possibly detected the brightest state (outburst)
of the source after around mid 1999 (previous optical outburst state). It is interesting that the 
interval between the last optical outburst in 1999 and the present detected possible outburst state 
is about 7.5 years, which is consistent with the prediction made by (Raiteri et al. 2006b). Our 
observations of the source was also the part of the late 2006 to March 2007 multi-wavelength campaign 
by WEBT. The multi-wavelength campaign results show the source was in the outburst state in this period 
(Raiteri et al. 2007b). 

\noindent
{\bf S5 0716$+$714}

S5 0716$+$714 is one of the brightest 
BL Lac objects. The optical continuum
is so featureless that all attempts made to determine its redshift have failed. The non detection
of host galaxy first set a lower limit of z $>$ 0.3 (Wagner et al. 1996) and very recently 
z $>$ 0.52 (Sbarufatti et al. 2005).  
This object is among one of the most well studied BL Lac objects for variability studies in the whole 
EM spectrum on diverse time scales (e.g. Wagner et al. 1990; Heidt \& Wagner 1996; Ghisellini et al. 
1997; Villata et al. 2000; Raiteri et al. 2003; Pian et al. 2005; Bach et al. 2005, 
2006; Nesci et al. 2005; Ostorero et al. 2006; Montagni et al. 2006; Foschini et 
al. 2006; Wu et al. 2007; Fan et al. 2007 and references therein). Wagner \& Witzel (1995) reported 
that the duty cycle of the source is 1 which implies that the source is almost always in the active 
state. First optical polarization study of this source was done by Takalo et al. (1994). They found that 
the source shows high polarization $\sim$ 20\% and intra-day polarization variability upto 3.5\%. 
Fan et al. (1997a) reported even much higher optical polarization up to 29\% in the source. 

Since 1994, people started observing this source intensely in optical bands. There are 4 major
optical outbursts reported in the source, these are: at the beginning of 1995, in late 1997, 
in the fall of 2001 and in March 2004 (Raiteri et al. 2003; Foschini et al. 2006). These 
4 outbursts give a possible period of long term variability of $\sim$ 3.0$\pm$0.3 years.  

We observed S5 0716$+$714 on January 10, 12; February 23; March 19 and 20; 2007  in R passband.
The light curves of the blazar and differential instrumental magnitude (star 5 - star 3)
with different arbitrary offset are displayed in different panels in Fig. 2. 
From Fig. 2, it is clear that the source has shown IDV on 2 nights (January 12 and February 23, 
2007) out of 5 nights of observations. We also performed the IDV detection test 
and found that the value of C for January 10, January 12, 
February 23, March 19 and March 20, 2007 are 1.80, 2.86, 3.88, 0.95 and 1.56 respectively, which 
also confirms that the source has shown IDV on January 12 and February 23. 
The variability amplitude on January 12 and February 23, 2007 are 6.3\% and 5.2\% respectively. 
Day to day variations are also distinctly visible in the Figs. 2 and 7. In Fig. 7, filled circles 
show the day average magnitude of S5 0716$+$714 and open circles show the day average differential 
magnitude of star 5 and star 3 with the same arbitrary constant added for all 5 observing nights.  
Fig. 7 shows clearly the day to day variations in the source and also shows the variability 
trend in the source. The source was at R $=$ 12.58 mag on January 12, 2007 which is comparable to 
the earlier outburst magnitude in R passband ($\sim$ 12.75 mag in beginning of 1995; 12.6 mag in 
late 1997; 12.55 mag in the fall of 2001) (Raiteri et al. 2003). So, we have possibly detected the 
brightest state of the outburst phase in the source on January 12, 2007. If the source has reached 
even brighter state than January 12, it will be before February 23, 2007. If we compare the detected 
possible outburst in the present observations, it gives a period of the present outburst is after 
$\sim$ 2.8 years from recent previous outburst in March 2004. It is consistent from the possible 
long term period of the source 3.0$\pm$0.3 years.
  
\noindent
{\bf PKS 0735$+$178}

PKS 0735$+$178 is an optically bright, highly variable source and classified as a BL Lac 
object (Carswell et al. 1974). 
The host galaxy of the source is unresolved in optical imaging. In the optical spectrum of 
PKS 0735$+$178, Carswell et al. (1974) did not find any emission line, but they noticed two 
sharp absorption features at 3981$\AA$ and 3991$\AA$ which they identified with the Mg II 
$\lambda$2798 doublet at redshift z $>$ 0.424 which sets a lower limit of its redshift. 
Very recently imaging redshift z = 0.424 is reported for the source using HST snapshot 
image (Sbarufatti et al. 2005). The 
source has been extensively studied in radio and optical bands in search for variability. 
In search for long term variability, in about a century long optical data using 
Jurkevich method, periodic components of 13.8$-$14.2 years were found (Fan et al. 1997b;
Qian \& Tao 2004). In a recent paper, Ciprini et al. (2007) has reported several short possible
periods. There are several attempts to search for optical IDV and day to day variation in the 
source in last about one and half decades. Optical IDV upto 0.5 mag has been reported in the 
source on some occasions (e.g. Xie et al. 1992; Massaro et al. 1995; Fan et al. 1997b; 
Zhang et al. 2004). 
PKS 0735$+$178 has shown high degree of optical and near-infrared polarizations which varies 
from 1\% to 30\% (e.g Mead et al. 1990; Takalo 1991; Takalo et al. 1992; Valtaoja et al. 1991, 
1993; Tommasi et al. 2001).        
    
We observed PKS 0735$+$178 on January 11, 2007 in R passband. The light curves of the blazar 
and differential instrumental magnitude (star C4  - star C1) with arbitrary offset are displayed 
in the second panel from the bottom in Fig. 6. From the figure it is clear that the source has 
shown IDV on one night of our observations. We also performed the IDV detection test 
and found that the value of C is 4.00, which also 
confirms that the source has shown IDV on one night of our observations. The observed variability 
amplitude is 4.9\%. Ciprini et al. (2007) have done the multi band optical monitoring of the source 
for more than ten years (1993 $-$ 2004) and also plotted its B passband historical light curve 
(1906 $-$ 2004). 1993 $-$ 2004 light curve of the source shows the brightest and faintest states 
of the source in R passband $\sim$ 14.0 and $\sim$ 17.0 magnitudes. In the present observations, 
we observed the source $\approx$ 1.7 magnitude fainter than the brightest state and $\approx$ 1.3 
magnitude brighter than its faintest state. 
Therefore, it appears likely that we have observed the source in pre/post outburst state.

\noindent
{\bf OJ 287}

Blazar OJ 287 at redshift z = 0.306 is one of the most extensively observed extragalactic 
object. In the long term optical light curve, it exhibits a $\sim$ 12 year periodic outburst 
with a double-peaked maxima in its flux variations (Takalo et al. 1994; Pursimo et al. 2000). 
Sillanp$\ddot{a}\ddot{a}$ et al. (1988) explained the periodic outburst in the blazar with a 
binary black hole model. In radio to optical bands, the source has been studied for flux and 
polarization variability on diverse time scales by several groups (e.g. Jorstad et al. 2007;
Hovatta et al. 2007; Zhang et al. 2007; Bach et al. 2007; Wu et al. 2006; Fan et al. 2006
and references therein). Takalo et al. (1994) have reviewed its observational properties from 
radio to X-ray bands.    

We observed OJ 287 on October 29, November 17, November 18, 2006; and  January 10, 2007 in 
R passband. The light curves of the blazar and differential instrumental magnitude (star 10 
- star 11) with same arbitrary offset are displayed in Fig. 3. 
From the figure it is clear that the source has not shown IDV on any of the four nights of our
observations. We also performed the IDV detection test 
and found that the value of C for October 29, November 17, November 18, 2006; 
and January 10, 2007 are 0.72, 1.00, 0.73 and 1.07 respectively, which also confirms that the 
source has not shown IDV on any of the 4 nights of our observations. In Fig. 7, filled circles 
show the day average magnitude of OJ 287 and open circles show the differential magnitude of 
star 10 and star 11 with the same arbitrary offset on all the 4 nights of observations. From 
Fig. 7, day to day variation in the source is clearly visible. Pursimo et al. (2000) have done 
the multi band optical monitoring of the source for about five years (1993 $-$ 1998) and also 
plotted its V passband historical light curve for about a century long observations. 1993 $-$ 
1998 light curve of the source shows the brightest and faintest states of the source in the 
R passband $\sim$ 13.5 and $\sim$ 16.5 magnitudes. In the present observations, we observed the 
source $\sim$ 14.6 $-$ 15.0 mag in R passaband which is $\approx$ 1.0 magnitude fainter than the 
brightest state and $\approx$ 1.5 magnitude brighter than its faintest state. 
Therefore, it appears likely that we have observed the source in pre/post outburst state. 
Our observations of the source were also the part of November
2006 multi-wavelength campaign by WEBT\footnote{Detail information of the WEBT campaign is 
available at http://www.to.astro.it/blazars/webt/}.  

\noindent
{\bf ON 231}

ON 231 at redshift z = 0.102 is classified as a BL Lac object (Browne 1971). Historical B passband 
light curve of the source (1935 - 1997) was plotted by Tosti et al. (1998). There are not many
attempts to study the variability behavior of the source. Source has shown optical flux variations 
on diverse time scales ranging from few hours to several years (Smith \& Nair 1995; Xie et al. 1992).       

We observed ON 231 on January 11, 2007 in R passband. The light curves of the blazar and 
differential instrumental magnitude (star D  - star A) with arbitrary offset are displayed 
in second panel from top in Fig. 6. From the figure it is clear that the source has not shown 
IDV on one night of our observations. We also performed the IDV detection test 
and found that the value of C is 1.06, which also confirms that the source has not shown IDV 
on one night of our observations. Tosti et al. (1998) 
have done the multi band optical monitoring of the source for three years (1994 $-$ 1997) in its
great outburst state. 1994 $-$ 1997 light curve of the source shows the brightest and faintest 
states in R passband $\approx$ 13.5 and $\approx$ 15.0 magnitudes. In the present observations, we 
observed the source at R $\sim$ 15.0 magnitude. 
Therefore, it appears likely that we have observed the source in the low-state
which is comparable to the faintest state observed by Tosti et al. (1998).  

\noindent
{\bf 3C 279}

3C 279 is an optical violent variable quasar and has shown large optical variation of $\Delta$B $\geq$
6.7 magnitude from archival plates of Harvard collection (Eachus \& Liller 1975). The most rapid 
optical variation $\Delta$V = 1.17 magnitude was reported in 40 min on May 22, 1996 (Xie et al. 1999). 
On another occasion the source brightened by $\sim$ 2.0 magnitude within 
24 hours (Webb et al. 1990). A large amplitude short term optical variability of $\Delta$ R = 0.91 
magnitude in 49 days (April to June 2001) has been reported earlier by (Xie et al. 2002b). It has 
shown very high optical polarization P = 43.3$\pm$1.3 \% (Scarpa \& Falomo 1997). The source has 
also shown optical polarization variability 17\% to 8\% in three nights and intra-day polarization 
is also detected (Andruchow et al. 2003). Using 27 years of near-infrared K band observations, 
Fan (1999) has reported that the source has shown 4.59 magnitude variations and a strong period 
of 7.1$\pm$0.44 years.  

We observed 3C 279 on January 12 and February 23, 2007  in R passband.
The light curves of the blazar and differential instrumental magnitude (star 8 - star 1)
with different arbitrary offset are displayed in Fig. 4. The top and bottom panels of Fig. 4 
show the observations on January 12 and February 23, 2007 respectively. 
From the figure it is clear that the source has not shown IDV on both the nights (January 12,
and February 23, 2007) of observations. We also performed the IDV detection test 
and found that the value of C for January 12 and February 23, 2007 are 1.0 and 1.0 respectively, 
which also confirms that the source has not
shown IDV on both the nights of our observations. 
Day to day variations are distinctly visible in the Fig. 4. In Fig. 7, filled circles show the day 
average magnitude of 3C 279 and open circles show the differential magnitude of star 8 and star 1
with same arbitrary offset on both the nights of observations. Fig. 7 also shows the variability 
trend and if the source has reached even brighter state than January 12, it was before February 
23, 2007. So, we have possibly observed the source in the outburst phase. We have detected large
amplitude $\sim \Delta$R = 1.5 magnitude variation in the source in the time span of 42 days
which is much larger variation with previous $\Delta$R = 0.91 magnitude in 49 days (April to June 
2001) reported by (Xie et al. 2002b). Our observations of the source was also the part of April 2006
$-$ February 2007 multi-wavelength observing campaign by WEBT\footnote{Detail information of the 
WEBT campaign is available at http://www.to.astro.it/blazars/webt/} in the outburst state of the 
source.

\noindent
{\bf 3C 454.3}

The blazar 3C 454.3 is a flat spectrum radio quasar at redshift z = 0.859. Long term 
optical and radio observations (1966 $-$ 2005) were displayed by Villata et al. (2006). 
They detected an unprecedented optical outburst in (2004 $-$ 2005) which lasted about a 
year in their multi wavelength observing campaign by WEBT. To further see the behavior 
of the source, two more multi-wavelength observing campaigns were organized by WEBT 
(Oct. 2005 $-$ June 2007; July 2007 - ongoing). Our observations of the source was also 
the part of Oct. 2005 $-$ June 2007 multi-wavelength observing campaign by WEBT.
 
We observed 3C 454.3 during Oct. 27 - 30, 2006 in V and R passbands and January 9-10, 
2007 in R passband. The light curves of the blazar and differential instrumental magnitude 
(star 3  - star 4) with same arbitrary offset for all 6 nights observations in R passband 
are plotted in the panels in the bottom row of the figure 5. V passband light curves for 
all 4 nights observations with same arbitrary offset are plotted in the panels in the top 
row of the figure 5. We performed the IDV detection test 
and found that the value of C in R passband 2.46, 2.64, 2.42, 
2.08, 3.83 and 1.77 for observations on October 27, October 28, October 29, October 30, 
2006; January 09 and January 10, 2007 respectively. The value of C shows that source has 
shown IDV in R passband on 2 nights. The variability amplitude on October 28, 2006 and 
January 09 2007 were 5.1\% and 17.1\% respectively. The value of C in V passband 2.05, 
1.79, 3.00 and 0.50 were obtained for observations on October 27, October 28, October 29 
and October 30, 2006 respectively. The value of C shows that source has shown IDV in V 
passband on one night. The variability amplitude on October 29, 2006 was 11.6\%. The present 
observations of the source was in its post outburst phase (Raiteri et al. 2007a).

\noindent
{\bf 1ES 2344+514}

1ES 2344+514 is classified as a BL Lac object based on lack of optical emission lines and 
CaII `break strength' less than 25\% (Perlman et al. 1996). It was first detected in TeV 
$\gamma-$rays ($>$ 350 GeV) by the Whipple group (Catanese et al. 1998) and discovered among 
one of the first five TeV detected blazars. First attempt of its optical IDV and short term 
variability was made by Xie et al. (2002a) in 6 observing nights during January 2000 to 
January 2001. They detect small amplitude IDV and short term variability in their observations.
  
We observed 1ES 2344+514 on January 12, 2007 in R passband. The light curves of the blazar and 
differential instrumental magnitude (star C2  - star C1) with arbitrary offset are displayed 
in the top panel in Fig. 6. From the figure it is clear that the source has not shown IDV on 
one night of our observations. We also performed the IDV detection test 
and found that the value of C is 1.44, which also 
confirms the non detection of IDV in source in one night of our observations. There is not 
much optical data available for the source, so, it is difficult to know in which state was 
the source during our observations. This source will remain one of our prime target 
to search for optical variability on diverse time scales in near future observations.  

\section {Discussion \& Conclusion}

From our observations, our data favor the hypothesis that three blazars: AO 0235$+$164, 
S5 0716$+$714 and 3C 279 were in the outburst state; one blazar: 3C 454.3 was in the 
post outburst state; three blazars: S2 0109$+$224, PKS 0735$+$178 and OJ 287 were in 
the pre/post outburst state; one blazar: ON 231 was in the low-state; and the state of 
one blazar: 1ES 2344$+$514 was not known because there is not much optical data available 
for the blazar. The available statistics is not sufficient to draw strong conclusions.

Detected outburst phase in AO 0235$+$164 possibly confirms the $\sim$ 8 years period 
in optical bands which was predicted by Raiteri et al. (2006b). AO 0235$+$164 has shown 
IDV on two night out of 3 nights of observations. One night in which we did not detected 
any IDV, observations duration were very short and no IDV could be detected (if it has 
really occurred on that night). IDV can be broadly divided into intrinsic and extrinsic 
classes. Extrinsic IDV is caused by refractive interstellar scintillation and only relevant 
in the low-frequency radio observations. Another extrinsic cause of IDV is gravitational 
micro-lensing. 
Optical imaging and spectroscopic observations of the source have revealed foreground 
absorbing systems at z = 0.524 and z = 0.851 (Cohen et al. 1987; Nilsson et al. 1996). 
The flux of the source is contaminated and absorbed by foreground absorbing systems, 
the stars of which can act as gravitational micro-lenses. The reported optical IDV in 
AO 0235$+$164 is well suited by the gravitational micro-lensing. 
Day to day variations are also noticed in our observations in the blazar. 

Out of 5 nights observations of S5 0716$+$714, IDV is detected in 2 nights of observations. 
Day to day variations are distinctly visible in all 5 nights of observations which confirms 
the variability detection on short time scale. Detected outburst possibly confirm the long 
term periodic variation time scale of 3.0$\pm$0.3 years. IDV reported in S5 0716$+$714 is 
intrinsic one. The dominantly fundamental model for intrinsic variability is shocks 
propagating through the jet (Blandford \& K$\ddot{o}$nigl 1979; Marscher \& Gear 1985). 
Models to explain intrinsic IDV are based on a relativistic shock propagating down a jet 
and interacting with irregularities in the flow (Marscher et al. 1992) or 
relativistic shocks changing directions in the jet (Nesci et al. 2005). Another related 
model involves non-axisymmetric bubbles carried outward in relativistic magnetized jets 
(Camenzind \& Krockenberger 1992). 
The IDV reported here in S5 0716$+$714 can be explained by any one of these models. 

Out of 2 nights observations of 3C 279, IDV is not detected. Day to day variations are shown 
in our two nights of observations of the source which confirms the variability detection on 
short time scale. We detected short term variation of $\sim$ 1.5 magnitude in R passband in
42 days which is much larger than the earlier detected short term variation of 0.91 magnitude
in 49 days (Xie et al. 2002b). When we observed 3C 279, the source was in outburst/high-state. 
In the densely sampled light curves in search for IDV in the source at outburst/high-state has 
not shown any IDV. No IDV detection in 3C 279 shows the jet emission do not have any 
irregularities in the jet flow, relativistic shocks directions have also not changed from 
the line of sight, no non-axisymmetric bubbles were carried outward in the relativistic 
magnetized jets.     

Two other blazars: PKS 0735$+$178 and 3C 454.3 have shown IDV in our observations. 3C 454.3 
was in the post outburst phase and PKS 0735$+$178 was in the pre/post outburst phase during 
our observations. The IDV detected in these two blazars is well suited for shock-in-jet models 
described above. In 3C 454.3, we have also noticed the short term variability. No IDV was 
detected in S2 0109$+$224, OJ 287, ON 231 and 1ES 2344$+$514 in our observations. In OJ 287,
we have noticed the short term variability.  

\acknowledgments

We thankfully acknowledge the critical comments by the referee which helped us to improve 
the paper significantly. We are thankful to Prof. W. Yuan for careful reading of the 
manuscript. ACG and JMB gratefully acknowledge the financial support from the 
national Natural Science Foundation of China (grant nos. NSFC 10533050 and NSFC 10573030). 
JHF's work is supported by the national Natural Science Foundation of China (grant nos. NSFC 
10573005 and NSFC 10633010).

\clearpage

\begin{deluxetable}{cccccccc}
\tabletypesize{\scriptsize}
\tablecaption{Complete log of optical photometric observations of nine blazars from 
1.02 meter Yunnan Astronomical Observatory Telescope, Kunming, China. 
\label{tbl-1}}
\tablewidth{0pt}
\tablehead{
\colhead{Blazar Name}    & \colhead{$\alpha_{2000.0}$} & \colhead{$\delta_{2000.0}$} & \colhead{z} 
& \colhead{Date of }     & \colhead{Filters}           & \colhead{Data} & \colhead{Duration}    \\      
                         &                             &                             &
& \colhead{Observations} &                             & \colhead{Points}     & \colhead{(hour)}   \\         
                         &                             &                             &
& \colhead{dd.mm.yyyy}   &                             &                   &                               
}
\startdata
S2 0109$+$224 & 01 12 05.7        & $+$22 44 39.2      & $\geq$0.4      & 11.01.2007 & R   & 80    & 3.00  \\
AO 0235$+$164 & 02 38 38.9        & $+$16 36 59.3      & 0.94        & 09.01.2007 & R   & 4     & 0.26  \\
              &                   &                    &             & 10.01.2007 & R   & 79    & 3.04 \\
              &                   &                    &             & 19.03.2007 & R   & 6     & 0.40  \\
S5 0716$+$714 & 07 21 53.5        & $+$71 20 36.4      & 0.3?, 0.52? & 10.01.2007 & R   & 116   & 3.03  \\
              &                   &                    &             & 12.01.2007 & R   & 185   & 3.56  \\
              &                   &                    &             & 23.02.2007 & R   & 60    & 1.00  \\
              &                   &                    &             & 19.03.2007 & R   & 82    & 1.58  \\
              &                   &                    &             & 20.03.2007 & R   & 195   & 4.24  \\
PKS 0735$+$178& 07 38 07.4        & $+$17 42 19.0      & 0.424       & 11.01.2007 & R   & 90    & 3.88  \\
OJ 287        & 08 54 48.9        & $+$20 06 31.0      & 0.306       & 29.10.2006 & R   & 15    & 1.14  \\
              &                   &                    &             & 17.11.2006 & R   & 18    & 1.18  \\
              &                   &                    &             & 18.11.2006 & R   & 16    & 1.06  \\
              &                   &                    &             & 10.01.2007 & R   & 49    & 1.79  \\
ON 231        & 12:21:31.7        & $+$28:13:58.5      & 0.102       & 11.01.2007 & R   & 72    & 3.24 \\
3C 279        & 12 56 11.2        & $-$05 47 21.5      & 0.5362      & 12.01.2007 & R   & 208   & 4.00  \\
              &                   &                    &             & 23.02.2007 & R   & 100   & 2.35  \\   
3C 454.3      & 22 53 57.7        & $+$16 08 54.0      & 0.859       & 27.10.2006 & R,V & 15,9  & 1.59 \\
              &                   &                    &             & 28.10.2006 & R,V & 15,8  & 1.27 \\ 
              &                   &                    &             & 29.10.2006 & R,V & 15,10 & 1.76 \\ 
              &                   &                    &             & 30.10.2006 & R,V & 14,2  & 1.10 \\ 
              &                   &                    &             & 09.01.2007 & R   & 11    & 0.68 \\ 
              &                   &                    &             & 10.01.2007 & R   & 7     & 0.54 \\ 
1ES 2344$+$514 & 23 47 04.8      & $+$51 42 17.9      & 0.044       & 12.01.2007 & R   & 53    & 1.89 \\
\enddata


%

\end{deluxetable}

\clearpage

\begin{deluxetable}{cccc}
\tabletypesize{\scriptsize}
\tablecaption{Standard stars in the observed blazars field.
\label{tbl-2}}
\tablewidth{0pt}
\tablehead{
\colhead{Blazar Name} & \colhead{Star} & \colhead{R magnitude} & \colhead{Reference$^{a}$} \\ 
                      & \colhead{No.}  & \colhead{(error)}     &  \\
}
\startdata
S2 0109$+$224 & I  & 12.11 (0.04) & 1  \\
              & C1 & 14.72 (0.06) & 1  \\
              & D  & 14.09 (0.05) & 1  \\
              & E  & 14.94 (0.05) & 1  \\
AO 0235$+$164 & 1  & 12.69 (0.02) & 2  \\
              & 2  & 12.23 (0.02) & 2  \\ 
              & 3  & 12.48 (0.03) & 2  \\ 
              & 6  & 13.64 (0.04) & 3  \\
              & C1 & 14.23 (0.05) & 3   \\
S5 0716$+$714 & 2  & 11.12 (0.01) & 4  \\
              & 3  & 12.06 (0.01) & 4  \\
              & 5  & 13.18 (0.01) & 4  \\
              & 6  & 13.26 (0.01) & 4  \\
              & 8  & 13.79 (0.02) & 4  \\
PKS 0735$+$178& C  & 13.85 (0.04) & 5  \\
              & C1 & 12.89 (0.04) & 5  \\
              & C2 & 12.79 (0.04) & 5  \\
              & C4 & 13.80 (0.04) & 5  \\
              & C7 & 14.70 (0.06) & 5  \\
OJ 287        & C1 & 15.50 (0.07) & 6  \\
              & C2 & 15.66 (0.08) & 6  \\
              & 4  & 13.74 (0.04) & 6  \\
              & 10 & 14.34 (0.05) & 6  \\
              & 11 & 14.65 (0.05) & 6  \\
ON 231        & A  & 11.72 (0.04) & 6  \\
              & C1 & 16.03 (0.10) & 6  \\
              & D  & 13.86 (0.04) & 6  \\
3C 279        & 1  & 12.05 (0.02) & 7 \\
              & 5  & 15.47 (0.02) & 7 \\ 
              & 8  & 14.44 (0.00) & 8 \\
3C 454.3      & 1  & 13.15 (0.02) & 7 \\
              & 2  & 13.19 (0.02) & 7 \\
              & 3  & 14.00 (0.02) & 7 \\
              & 4  & 14.79 (0.02) & 7 \\
              & 5  & 14.83 (0.02) & 7 \\
              & 6  & 14.83 (0.03) & 7 \\
              & 7  & 14.94 (0.02) & 7 \\
              & 8  & 15.34 (0.02) & 7 \\
1ES 2344$+$514 & C1 & 12.25 (0.04) & 3 \\
                & C2 & 14.20 (0.05) & 3 \\
                & C2 & 15.40 (0.08) & 3 \\ \hline
Blazar        & Star & V magnitude  & Reference$^{a}$ \\
              & No.  & (error)      &    \\ \hline
3C 454.3      & 1  & 13.71 (0.02) & 7 \\
              & 2  & 13.80 (0.02) & 7 \\
              & 3  & 14.44 (0.02) & 7 \\
              & 4  & 15.21 (0.02) & 7 \\
              & 5  & 15.30 (0.02) & 7 \\
              & 6  & 15.34 (0.03) & 7 \\
              & 7  & 15.74 (0.02) & 7 \\
              & 8  & 15.94 (0.02) & 7 \\
\enddata


%
\tablenotetext{a}{1. Ciprini et al. 2003; 2. Smith et al. 1985; 3. Fiorucci et al. 1998; 4. Villata et al.
1998; 5. Ciprini et al. 2007; 6. Fiorucci et al. 1996; 7. Raiteri et al. 1998; 9. Villata et al. 1997}

\end{deluxetable}

\clearpage

\begin{deluxetable}{ccccccccccc}
\tabletypesize{\scriptsize}
\tablecaption{Results of IDV observations of the blazars. V
and NV in the Variable column represent variable
and non variable respectively. N represent the number of data points.
\label{tbl-3}}
\tablewidth{0pt}
\tablehead{
\colhead{Date}       & \colhead{Blazar} & \colhead{Time Range} & \colhead{Filter} & \colhead{N} & \colhead{$\sigma_{(BL - S_{A})}$} 
& \colhead {$\sigma_{(BL - S_{B})}$} & \colhead{$\sigma_{(S_{A} - S_{B})}$}  & \colhead{Variable} & \colhead{C} & \colhead{A\%} \\
\colhead{dd.mm.yyyy} & \colhead{Name}   & \colhead{UT}   &      &             &              & 
&  &                    &             &
}
\startdata
11.01.2007 & S2 0109$+$224 & 11.36 $-$ 14.35 & R & ~80 & 0.007    & 0.008  & 0.007 & NV & 1.07 & -- \\
09.01.2007 & AO 0235$+$164 & 12.60 $-$ 12.86 & R & ~~4 & 0.007    & 0.009  & 0.004 & NV & 2.00 & --  \\
10.01.2007 &               & 13.10 $-$ 16.14 & R & ~79 & 0.029    & 0.029  & 0.007 & ~V & 4.14 & 13.7  \\
19.03.2007 &               & 11.96 $-$ 12.36 & R & ~~6 & 0.040    & 0.042  & 0.011 & ~V & 3.73 & ~9.5  \\
10.01.2007 & S5 0716$+$714 & 16.34 $-$ 19.37 & R & 116 & 0.009    & 0.009  & 0.005 & NV & 1.80 & --  \\
12.01.2007 &               & 13.92 $-$ 17.71 & R & 185 & 0.019    & 0.021  & 0.007 & ~V & 2.86 & 6.3  \\
23.02.2007 &               & 12.22 $-$ 13.22 & R & ~60 & 0.016    & 0.015  & 0.004 & ~V & 3.88 & ~5.2  \\
19.03.2007 &               & 12.60 $-$ 14.18 & R & ~82 & 0.006    & 0.013  & 0.010 & NV & 0.95 & --  \\
20.03.2007 &               & 11.99 $-$ 16.25 & R & 195 & 0.023    & 0.027  & 0.016 & NV & 1.56 & --  \\
11.01.2007 & PKS 0735$+$178& 14.57 $-$ 18.45 & R & ~90 & 0.012    & 0.012  & 0.003 & ~V & 4.00 & ~4.9 \\
29.10.2006 & OJ 287        & 19.58 $-$ 20.72 & R & ~15 & 0.027    & 0.044  & 0.049 & NV & 0.72 & --    \\
17.11.2006 &               & 21.58 $-$ 22.76 & R & ~18 & 0.006    & 0.008  & 0.007 & NV & 1.00 & --  \\
18.11.2006 &               & 21.69 $-$ 22.75 & R & ~16 & 0.013    & 0.028  & 0.028 & NV & 0.73 & --  \\
10.01.2007 &               & 19.61 $-$ 21.40 & R & ~49 & 0.016    & 0.016  & 0.015 & NV & 1.07 & --  \\
11.01.2007 & ON 231        & 18.93 $-$ 22.17 & R & ~72 & 0.010    & 0.009  & 0.009 & NV & 1.06 & -- \\
12.01.2007 & 3C 279        & 18.96 $-$ 22.96 & R & 208 & 0.008    & 0.010  & 0.009 & NV & 1.00 & --  \\
23.02.2007 &               & 17.82 $-$ 20.17 & R & 100 & 0.005    & 0.009  & 0.007 & NV & 1.00 & --  \\
27.10.2006 & 3C 454.3      & 11.76 $-$ 13.10 & R & ~15 & 0.036    & 0.028  & 0.013 & NV & 2.46 & --  \\
           &               & 12.05 $-$ 13.35 & V & ~~9 & 0.022    & 0.023  & 0.011 & NV & 2.05 & --  \\
28.10.2006 &               & 15.06 $-$ 16.33 & R & ~15 & 0.018    & 0.019  & 0.007 & ~V & 2.64 & ~5.1 \\
           &               & 15.44 $-$ 16.11 & V & ~~8 & 0.041    & 0.034  & 0.021 & NV & 1.79 & --  \\
29.10.2006 &               & 11.15 $-$ 12.54 & R & ~15 & 0.014    & 0.015  & 0.006 & NV & 2.42 & --  \\
           &               & 11.44 $-$ 12.90 & V & ~10 & 0.042    & 0.036  & 0.013 & ~V & 3.00 & 11.6 \\
30.10.2006 &               & 11.56 $-$ 12.66 & R & ~14 & 0.013    & 0.012  & 0.006 & NV & 2.08 & --  \\
           &               & 11.86 $-$ 11.95 & V & ~~2 & 0.013    & 0.000  & 0.013 & NV & 0.50 & --  \\
09.01.2007 &               & 11.62 $-$ 12.30 & R & ~11 & 0.059    & 0.056  & 0.015 & ~V & 3.83 & 17.1 \\
10.01.2007 &               & 11.95 $-$ 12.49 & R & ~~7 & 0.054    & 0.077  & 0.037 & NV & 1.77 & --  \\
12.01.2007 & 1ES 2344$+$514 & 11.61 $-$ 13.50 & R & ~53 & 0.010  & 0.013  & 0.008 & NV & 1.44 & --  \\
\enddata
\end{deluxetable}

\clearpage

\begin{deluxetable}{cc}
\tabletypesize{\scriptsize}
\tablecaption{R passband Photometric data of observed blazars. Complete table will be publish
electronically.
\label{tbl-4}}
\tablewidth{0pt}
\tablehead{
}
\startdata
Blazar Name & Date \\
S2 0109$+$224 & 11.01.2007 \\\hline
UT            & Magnitude \\\hline
   11.35889 &  14.148  \\
   11.39917 &  14.155\\
   11.43583 &  14.142\\
   11.47250 &  14.145\\
   11.51083 &  14.139\\
   11.54833 &  14.144\\
   11.58556 &  14.138\\
   11.62194 &  14.140\\
   11.65917 &  14.140\\
   11.69639 &  14.144\\  
\enddata

\end{deluxetable}

\clearpage

\begin{figure}
\plotone{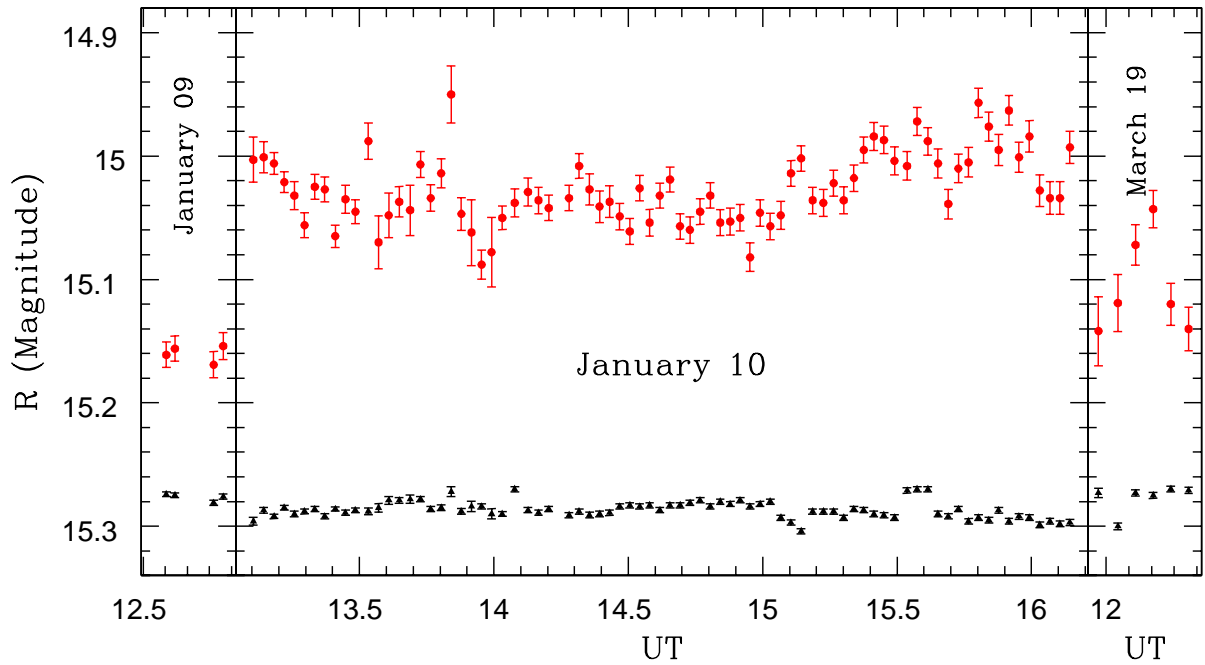}
\caption{R band light curve of AO 0235$+$164 (filled circles) and differential
instrumental magnitude of standard stars (Star1$-$Star3) (filled triangles) on
the nights January 9, January 10 and March 19, 2007 (from left to right panels) 
respectively. Standard stars differential light curve is offset for clarity by same 
arbitrary constant on all the three nights of observations.}
\end{figure}

\begin{figure}
\plotone{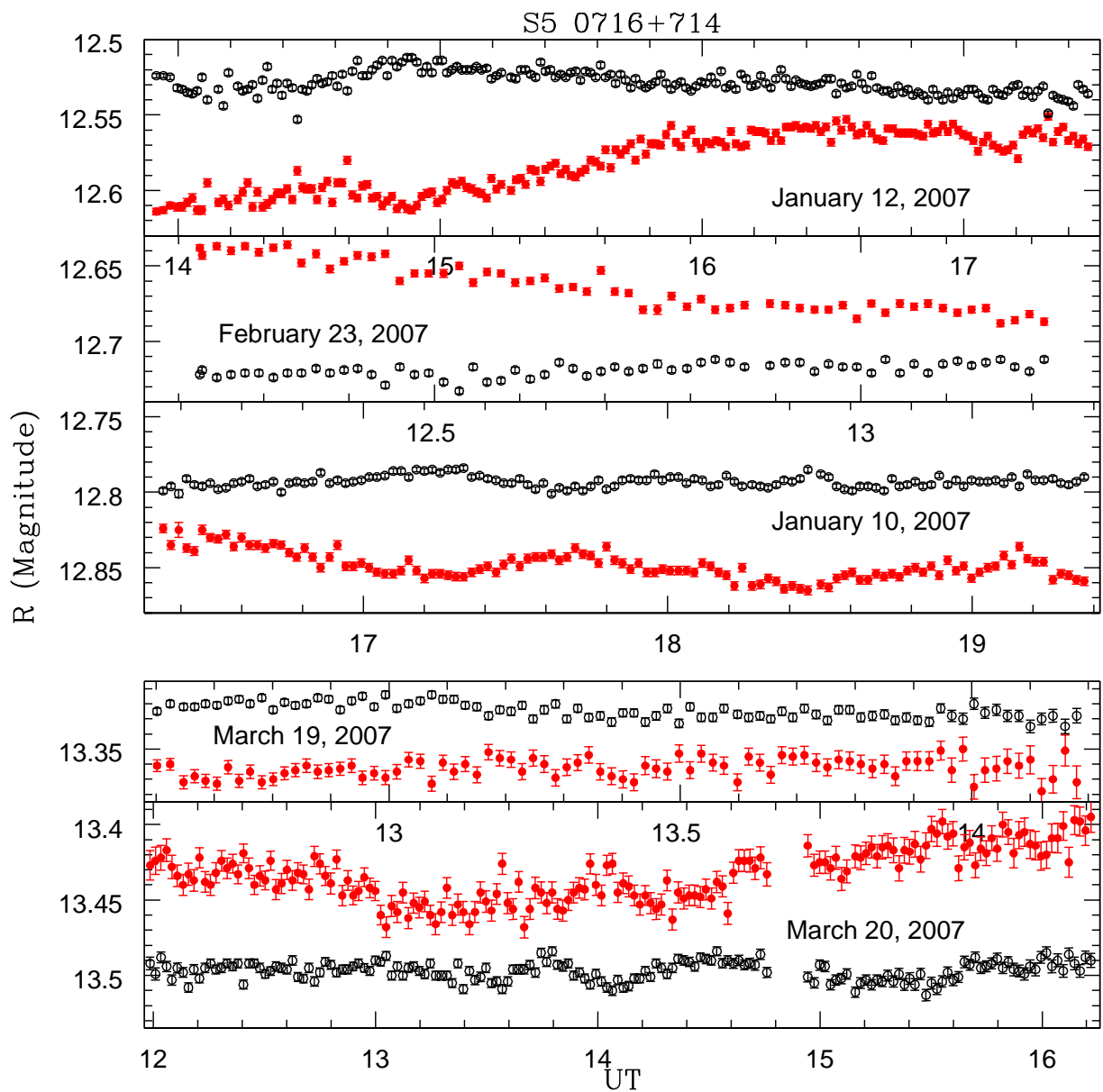}
\caption{R band light curve of S5 0716$+$714 (filled circles) and differential
instrumental magnitude of standard stars (Star5$-$Star3) (open circles) on
the nights January 12, February 23, January 10, March 19 and March 20 2007 (from top
to bottom panels) respectively. Standard stars differential light curve is offset
for clarity by different arbitrary constants on the 5 nights of observations.}
\end{figure}

\begin{figure}
\plotone{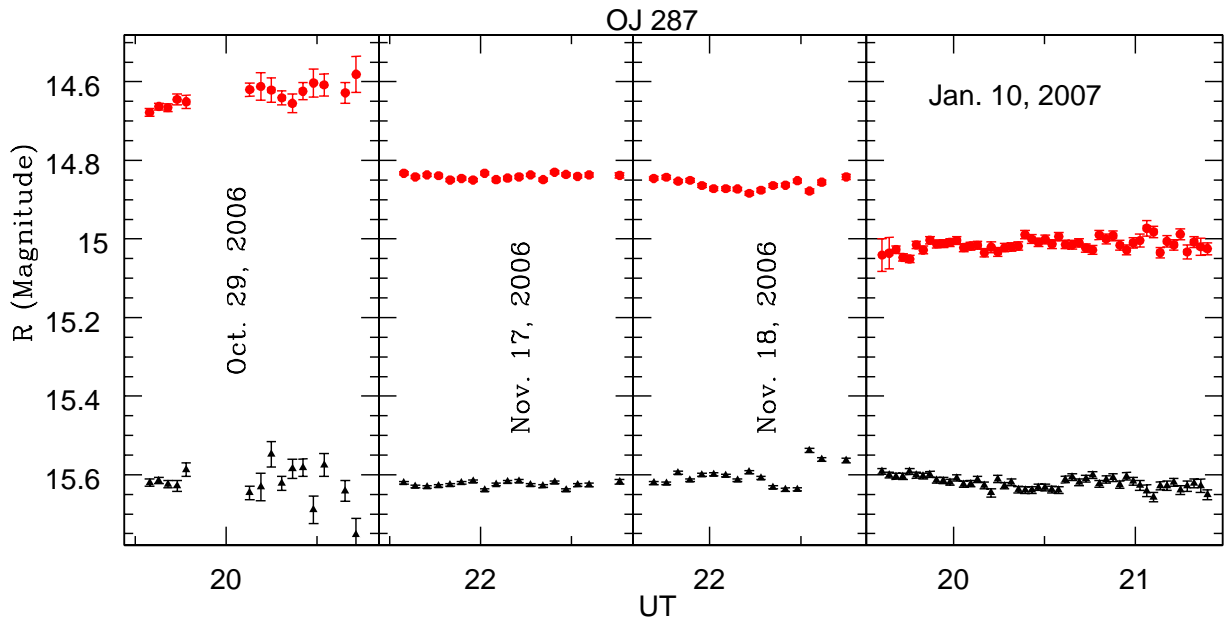}
\caption{R band light curve of OJ 287 (filled circles) and differential instrumental 
magnitude of standard stars (Star10$-$Star11) (filled triangles) on the nights October 29,
November 17, November 18, 2006 and January 10, 2007 (from left to right panels) respectively. 
Standard stars differential light curve is offset for clarity by same arbitrary constant 
on the all 4 nights of observations.}
\end{figure}

\begin{figure}
\plotone{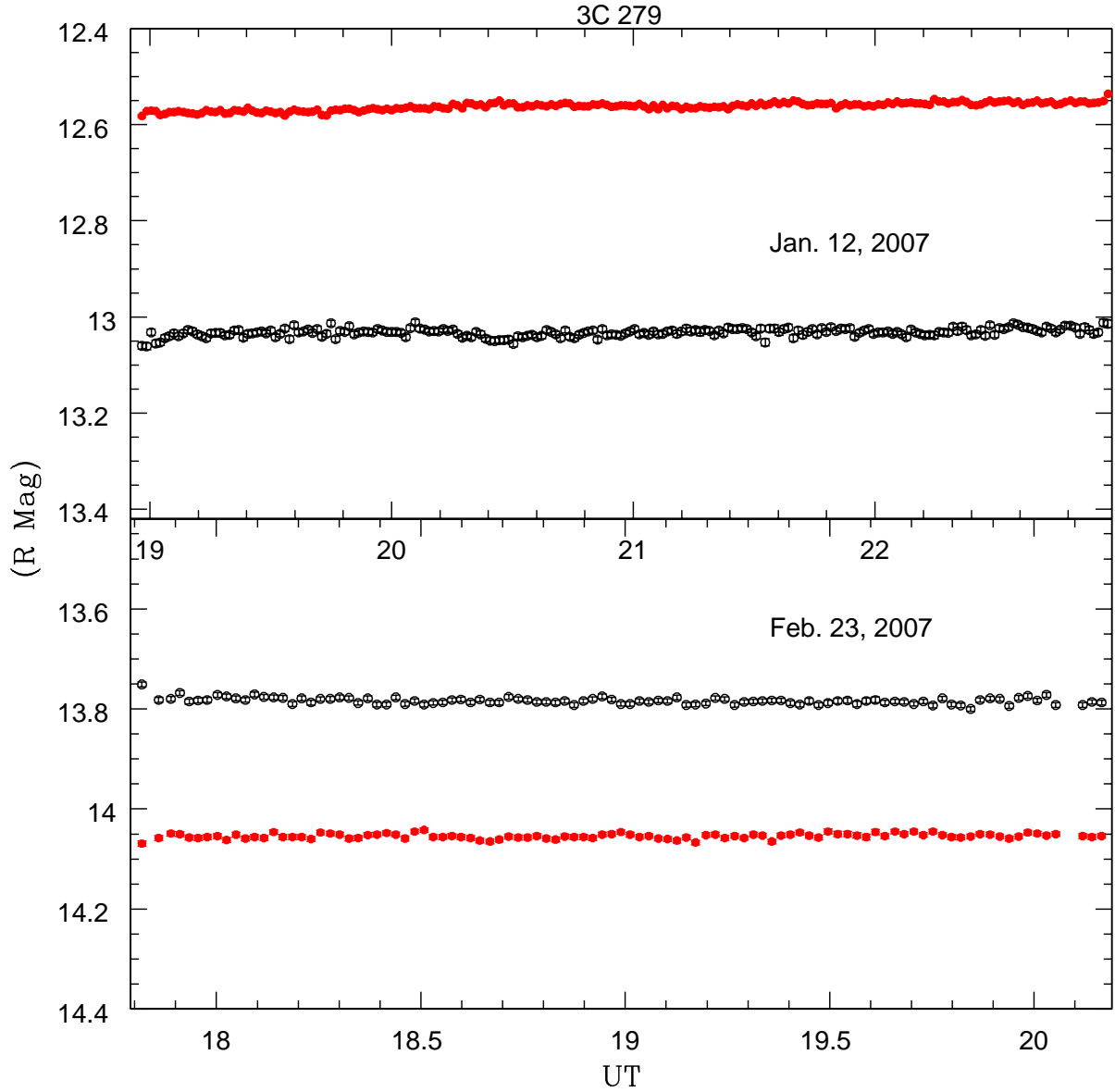}
\caption{R band light curve of 3C 279 (filled circles) and differential
instrumental magnitude of standard stars (Star8$-$Star1) (open circles) on
the nights of January 12 (top panel) and February 23, 2007 (bottom panel). Standard stars 
differential light curve is offset for clarity by different arbitrary constants on both 
the nights of observations.}
\end{figure}

\begin{figure}
\plotone{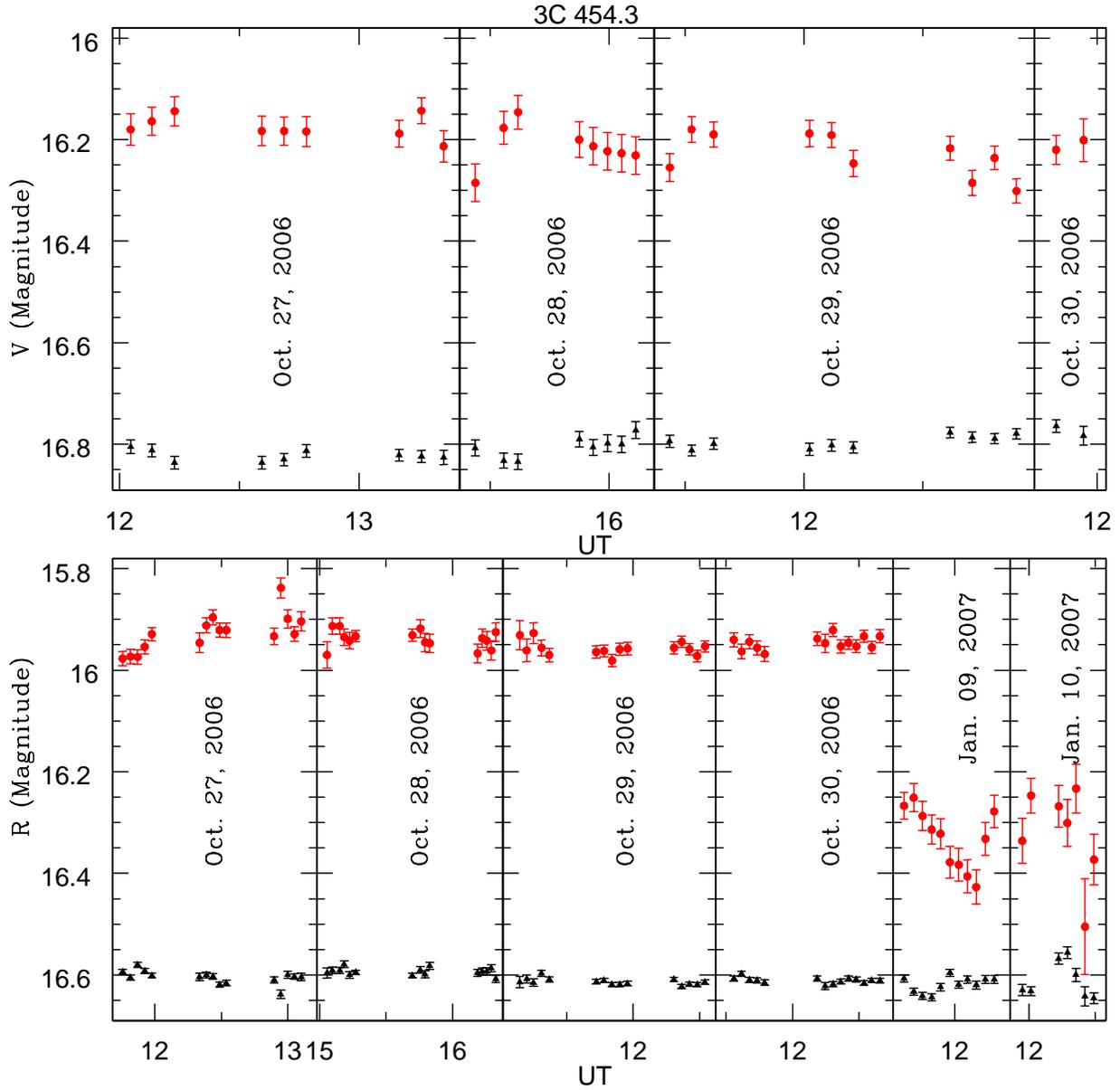}
\caption{V and R bands light curve of 3C 454.3 (filled circles) and differential
instrumental magnitude of standard stars (Star3$-$Star4) (filled triangles) on
panels in top row and panels in bottom row respectively. Date of observations are
marked in individual panels. Standard stars differential light curve is offset for 
clarity by the one arbitrary constants for R and another arbitrary constants for V 
on all the nights of observations.}
\end{figure}

\begin{figure}
\plotone{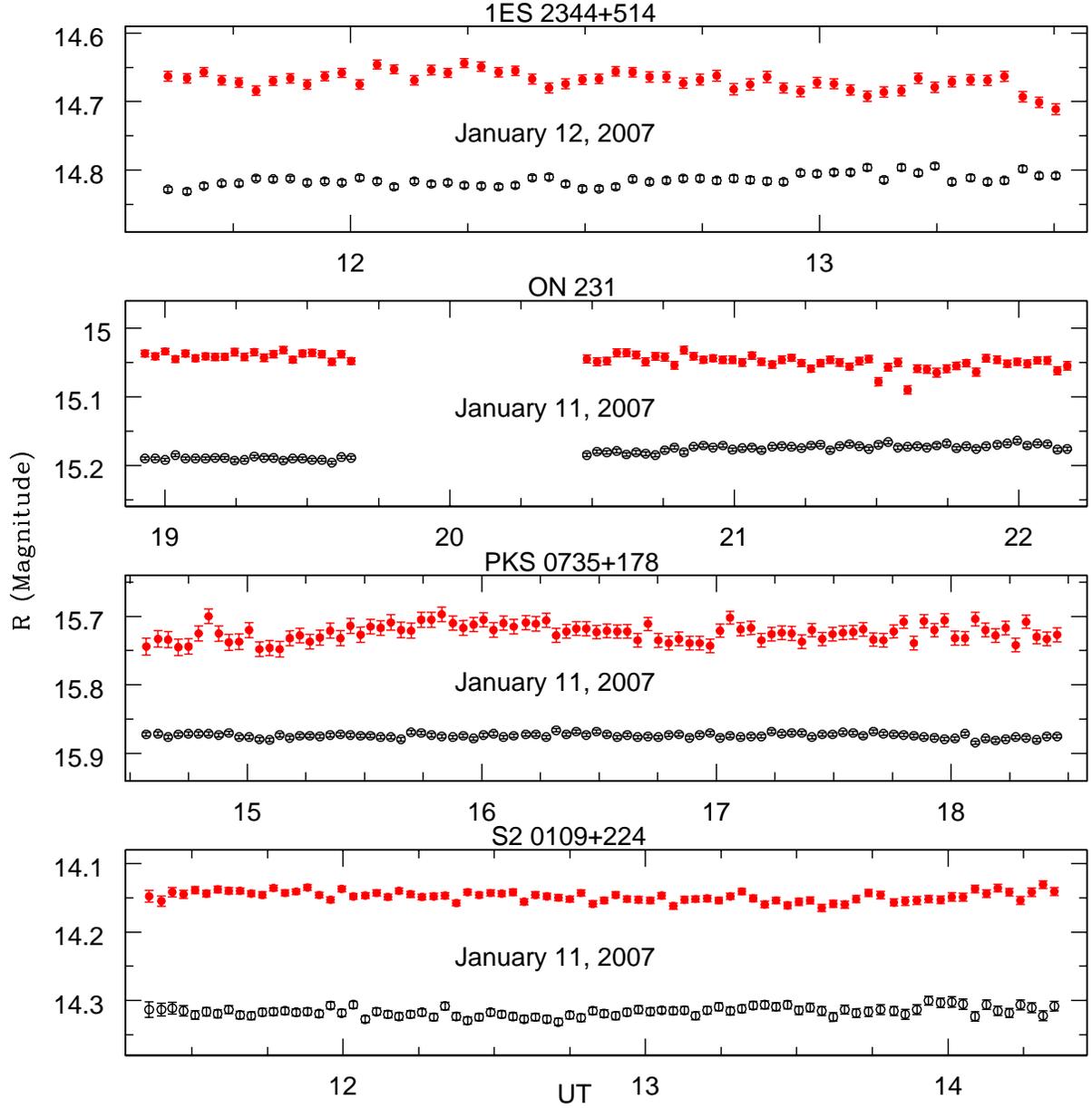}
\caption{R band light curves of S2 0109$+$224, PKS 0735$+$178, ON 231 and 1ES 2344$+$514 
(filled circles) from bottom to top panels respectively. For S2 0109$+$224, differential 
instrumental magnitude of standard stars (Star D $-$ Star C1); for PKS 0735$+$178,
differential instrumental magnitude of standard stars (Star C4 $-$ Star C1); for ON 231,
differential instrumental magnitude of standard stars (Star D $-$ Star A); for
1ES 2344$+$514, differential instrumental magnitude of standard stars (Star C2 $-$ Star C1)  
are plotted by (open circles) for clarity by using arbitrary constants.}
\end{figure}

\begin{figure}
\plotone{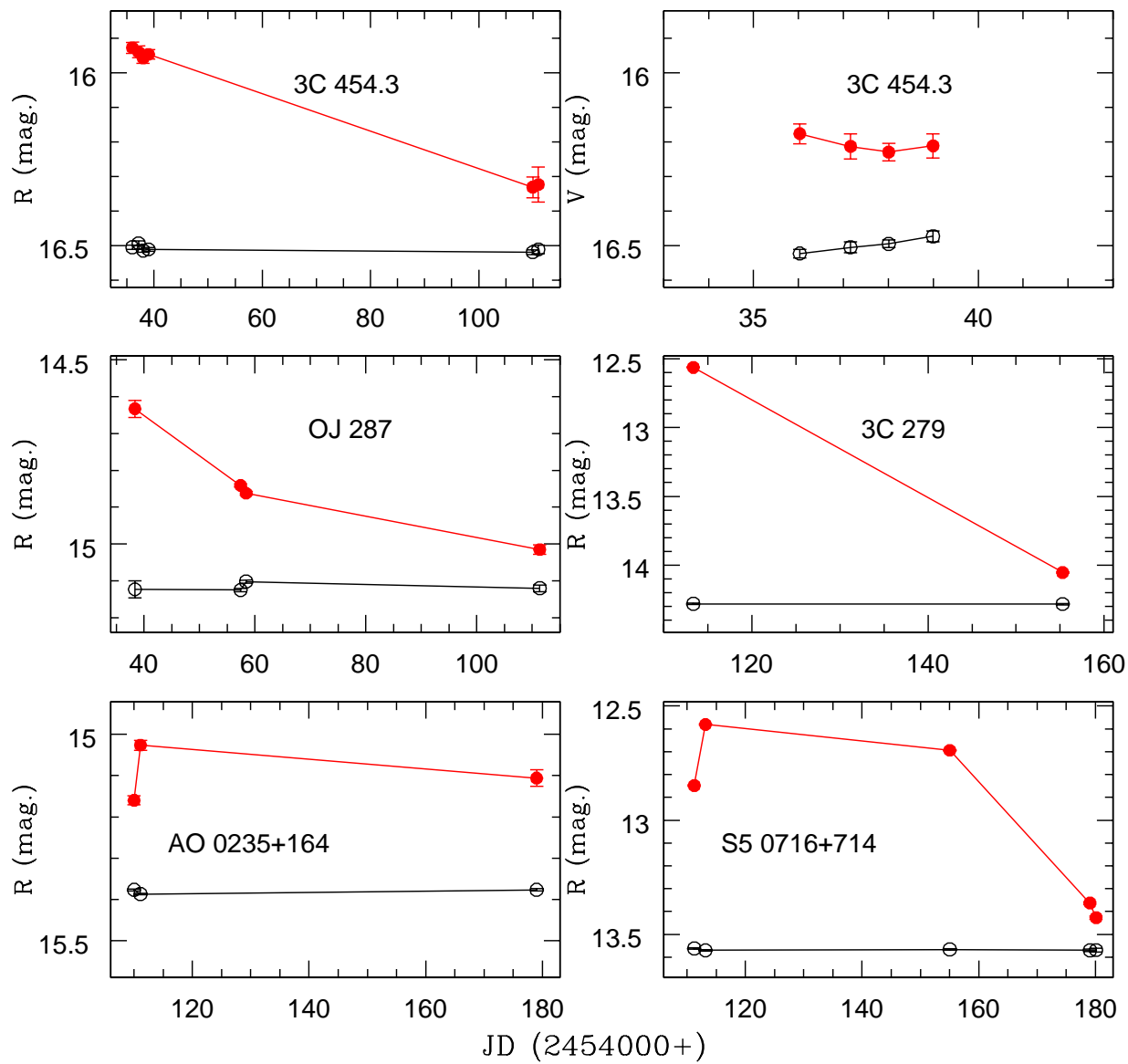}
\caption{The figure show the average R band magnitudes of the blazars AO 0235$+$164, 
S5 0716$+$714, OJ 287, 3C 279 and 3C 454.3 (between October 27, 2006 to March 20, 2007). 
Top left panel of the figure shows the average V band magnitude of the blazar 3C 454.3 in 
4 nights (October 27-30, 2006). Average magnitude of blazars are shown by filled circles. 
Open circles show the differential instrumental magnitudes of two standard stars in the 
blazar field offset by same arbitrary constant for specific blazar on all observing 
nights for clarity.}
\end{figure}

\end{document}